\documentstyle[11pt,newpasp,twoside,epsf]{article}
\begin{document}
\title{Are peculiar Wolf-Rayet Stars of type WN8 Thorne-$\dot{\mbox{Z}}$ytkow
Objects?}
\author{C\'edric Foellmi \& Anthony F.J. Moffat}
\affil{D\'epartement de physique, Universit\'e de Montr\'eal, c.p. 6128, succ. 
    centre-ville, Montr\'eal (Qu\'ebec) H3C 3J7, Canada}

\begin{abstract}
Most population I Wolf-Rayet (WR) stars are the He-rich descendants of
the most massive stars ($M_{i} = 25 - 100 M_{\odot}$).  Evidence has been
accumulating over the years that among all pop I WR stars, those of the
relatively cool, N-rich subtype "WN8" are among the most peculiar: \\
1. They tend to be runaways, with large space velocity and/or avoid
clusters. 2. Unlike their equally luminous WN6,7 cousins, only a very small
number of WN8 stars are known to belong to a close binary with an OB companion. 
3. They are the systematically most highly stochastically variable among
all (single) WR stars. \\
Taken together, these suggest that many WN8 stars may originally have
been in close binaries (like half of all stars), in which the original
primary exploded as a supernova, leaving behind a very close binary
containing a massive star with a neutron star/black hole companion (like
Cyg X-3).  When the massive remaining star evolved in turn, it engulfed
and eventually swallowed the compact companion, leading to the presently
puffed-up, variable WN8 star.  Such stars could fall in the realm of
the exotic Thorne-$\dot{\mbox{Z}}$ytkow objects.
\end{abstract}

\section{Introduction}
Thorne-$\dot{\mbox{Z}}$ytkow Objects (TZOs) have been proposed for the first time by Thorne
and $\dot{\mbox{Z}}$ytkow (1977). They are stars with a degenerate neutron core which
provides peculiar conditions to ensure the presence of a nuclear burning
region. This supports the envelope of the star, which appears like a Red Supergiant 
(see e.g. Biehle 1991, 1994; Cannon et al. 1992; Cannon 1993). Their lifetime in
the RSG phase is also expected to last as long as for a normal RSG (i.e. with 
nuclear burning right to the center of the core). Being RSGs (or appearing so), 
TZOs probably have strong winds.

We develop here the idea that if TZOs evolve like normal RSGs do, the most
massive of them will evolve further to the WR stage, following the usually
adopted scenario that RSG become WR stars because of strong mass loss by
stellar wind, stripping the outer envelope (see e.g. Garcia-Segura et al. 1996).

Among the population I Wolf-Rayet stars, those of subtype WN8 are peculiar.
We present here why they are good candidates to be "evolved" TZOs, 
formed via a binary scenario.

This paper is organized as follows: in section 2 we give the details of two
binary scenarii following which TZOs can be formed. These two scenarii
provide the same paradigm to explain TZOs and objects like Cyg X-3. 
In section 3 we discuss how these WN8 stars
are really different and how they can really come from a binary
evolution scenario. In section 4 we discuss observations which could be
undertaken to confirm such exotic objects. Section 5 gives our conclusions.

\section{Two scenarii to form TZO}

In fact three scenarii to form TZOs appear in the literature. 
In two of them, which are discussed below, they are the result of massive 
close binary evolution. The point here is that they also provide a reasonable
explanation for the characteristics of WN8 stars. The third scenario is
discussed by Davies, Benz \& Hills (1992) and will not be repeated here; it concerns 
the collision of a neutron star 
with a massive main sequence star in a star cluster. 

The two scenarii discussed below are roughly the same. TZOs
can be formed in a massive close binary in which the secondary goes through a
RSG phase, during which it engulfs the neutron star, arising from the normal
evolution of the primary. In order to form TZOs, the orbital period of the 
system is crucial, as well as the masses of the two components. Terman,
Taam \& Hernquist (1995, hereafter TTH95) have performed hydrodynamical 
simulations of a neutron star
entering the envelope of a more-or-less evolved RSG. Their results can be
used to construct one paradigm explaining both the origin of evolved TZO 
and objects like Cyg X-3.

\subsection{First scenario}

In the first scenario, the binary system is composed of an O-star and a
Wolf-Rayet star, the latter of which is the primary (i.e. the initially more massive star).
At the end of its evolution, following the usually adopted scenario of a
WR star (Chiosi \& Maeder 1986), the primary explodes as a supernova (SN) 
leaving behind a newly formed neutron star (NS). The rapid SN mass-loss by the
primary gives a kick velocity to the system ranging from 30 to $\sim$450 km/s
and more (De Donder, Vanbeveren, \& Van Bever 1997). 

Then, the system is composed of an O-star and a NS. If the secondary
has an initial mass between 25 and 40$M_{\odot}$, the evolutionary
sequence follows: O$\rightarrow$ Of$\rightarrow$ RSG$\rightarrow$ WN8$\rightarrow$ 
WNE$\rightarrow$ WC$\rightarrow$ SN (Crowther et al. 1995).
Although the orbital period has certainly changed from what it was 
originally, it is certainly feasible in some cases that the period 
could be sufficiently short to allow common envelope evolution. During the
growth of the secondary envelope towards a RSG, a gradual
coalescence of the NS and the secondary occurs. The NS then spirals in to the core of
the RSG, which loses is loosing a significant fraction of its envelope (see
below).

\subsection{Second scenario}

In the second scenario, the explosion of the primary
is asymmetric. The amplitude and the direction of kick place the newly
formed NS into a bound orbit with a periastron distance smaller than the radius 
of the secondary (Leonard et al. 1994). This process
depends obviously on the orbital period. In that case 
the secondary must already be a RSG (implying an initial mass ratio closer to unity, 
so the evolutionary timescales of both stars are comparable), to ensure a
non-negligible cross section. For a detailed description of how a SN
explosion affects the orbital characteristics in a close binary, see, e.g.
Kalogera (1996).

To first approximation we can suppose
that the spiral-in of the NS inside the envelope of the secondary follows the
same evolution as in the first scenario, apart from possible differences in
timescales.

\subsection{The common envelope phase}

The spiral-in process lasts only a few years (TTH95). Although some
authors (Armitage \& Livio 2000) argued that a newly formed TZO experiences dramatic
neutrino-losses (which implies a collapse of the object into a BH soon after
its formation), other hydrodynamical simulations (e.g. Biehle 1991) show
the presence of a tiny burning region ($\sim$ 40 meters!) just above the
degenerate neutron core which sustains the envelope mainly via 
rapid-proton burning. This process stops when half of the fuel (heavy
elements) is burned; the lifetime of the TZO RSG is as long as it is for a
"normal" RSG (Biehle 1991). 

Following the simulations
of Terman et al. (1995), it seems important that the secondary be at the 
beginning of the RSG phase to ensure that only \textit{a fraction} of its
envelope is lost. In fact, this fraction depends directly on the density
gradient of the RSG envelope. In the case of a "young" RSG, this gradient
is steep, and a large part of the mass of the star is located close to the
core. When a NS spirals inside the envelope, $\sim$ 20\% of the
mass of the envelope is lost (sequence 1 of TTH95). This means that
the NS stays inside the RSG and does not reappear later on. We are faced
with  two possibilities:
either this fraction is sufficient for the system to appear like a WR star,
or it is not sufficient, so that the system looks like a RSG and only a stellar
wind will be able to remove enough of the envelope to allow visible
caracteristics of a WR star to appear.

\subsection{Cyg X-3: a "failed" TZO?}

Cyg X-3 is a well known binary system, and a strong modulated source of X-rays.
It contains a Wolf-Rayet star of subtype probably WN7 and a NS orbiting with a period
of only 4.8 hours (see e.g. Cherepashchuk \& Moffat 1994). Within the same 
paradigm, the binary evolution scenario with common envelope evolution
described above can lead to the formation of such an object. (In fact, the first mention
about a possible link between TZOs, WN8 stars and Cyg X-3 was proposed by 
Cherepashchuk \& Moffat 1994).

Indeed, if the secondary was evolved, 
the density gradient inside the envelope would be flatter, and between 85
and 100 \% of the RSG envelope is lost (TTH95). In that case, the NS reappears, and
the system acquires a period of only a few hours.

More generally, there exists a correlation between the initial period of the
system (i.e. when the binary system is O/RSG + WR) and the possible
occurence of the common envelope evolution, leaving a TZO or a "Cyg
X-3" -type object (see Fig. 1).

\begin{figure}
\plotone{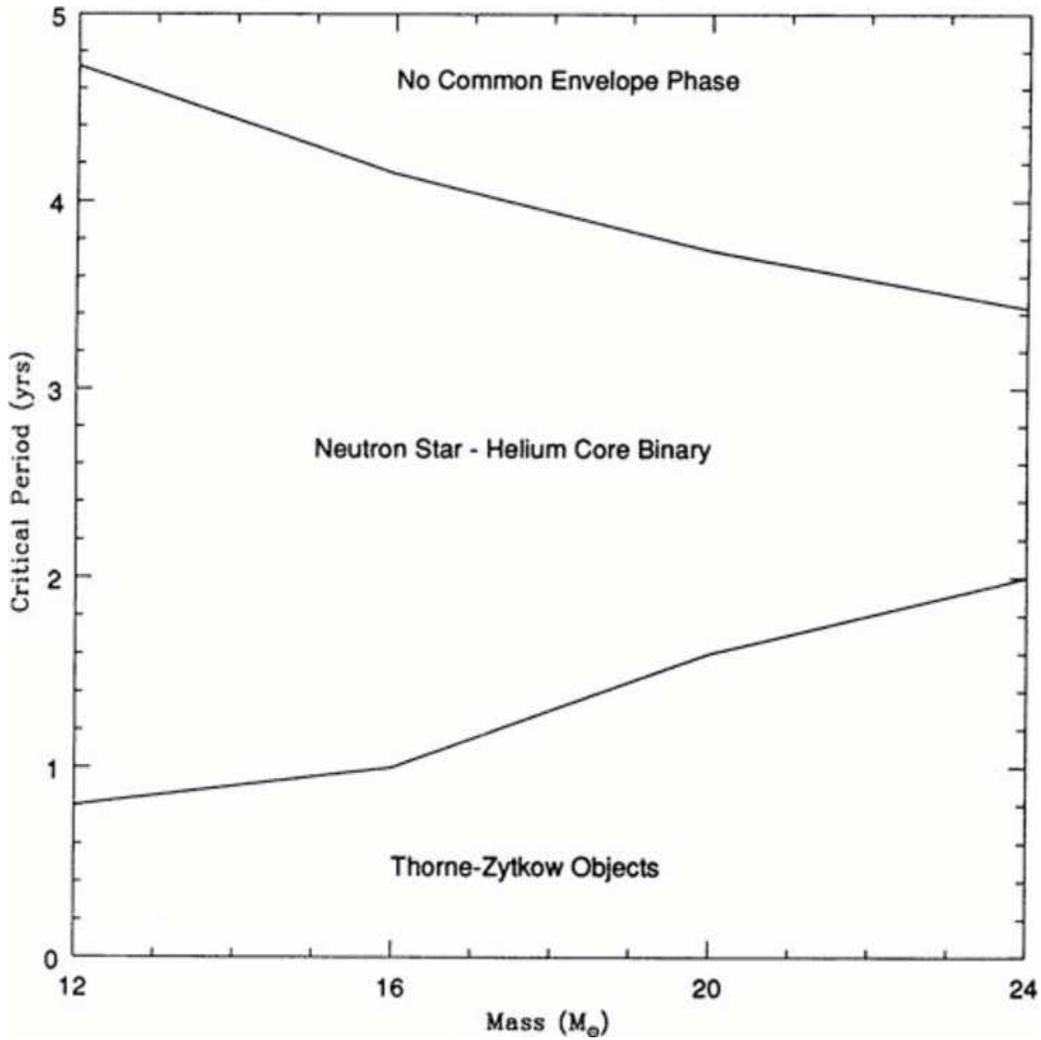}
\caption{Critical period as a function of the mass of the primary (in units
of $M_{\odot}$). Figure from TTH95. The initial period of Cyg X-3 should lie
in the intermediate region.}
\end{figure}

\section{Why WN8 stars?}

Among the population I Wolf-Rayet stars, WN8 stars show striking
peculiarities. They possess three main
caracteristics which differ from "normal" WR stars:
\begin{enumerate}
\item They have often a large space velocity and/or they avoid clusters and
associations (Moffat 1989). Most of the single WN8 stars in the Galaxy show
a large distance from the Galactic plane (van der Hucht 2000).
\item Only a very few known WN8 stars in the Galaxy and the Large Magellanic
Cloud (LMC) belong to a massive WR+O binary (Moffat 1989; van der Hucht 2000).
\item They appear to be stochastically variable, as much  as polarimetrically
photometrically (Drissen et al. 1987; Marchenko et al. 1998).
\end{enumerate}

In order to explain these three features with one paradigm, we propose
that WN8 stars are the result of massive close binary evolution in which the
primary has exploded as a SN, became a NS and spiraled in to the center of
the secondary. Indeed, the kick velocity often observed (or the large
distance to the clusters/associations or Galactic plane) is provided by the
explosion of the primary. If the NS spirals inside the envelope of the
secondary and remains within (i.e. only a fraction of the envelope of the RSG
is lost), it is evident that the system will not exhibit binary characteristics
(no radial velocity variations can be detected).

An important point to note here is the following: Marchenko et al. (1998) showed, 
based on a statistical trend, that the variability observed in WN8 stars is due to
core activity. Can we then ask: Could the presence of a NS in the core (even
slightly off-centered) be responsible for instabilities which affect the 
behaviour of the envelope and then the observational characteristics? 
    
\section{Can TZOs be observed?}

How can we detect and observe TZOs? Indeed, as first proposed by Cannon
(1992), the burning process is a rapid-proton chain, which produces a different 
set of heavy elements from what we can observe in normal stars. 
Usually these elements are partially brought
up to the surface, via convection inside the envelope. If the burning process 
supporting the envelope is due to
the presence of a degenerate neutron core, and is located in a tiny shell
above the core, unusual heavy elements, such as Mo should be
created (Biehle 1994).

But if the mass loss due to the spiral-in process is not
sufficient for the secondary to look like a WN8 star, the envelope of the RSG will
be large and it would be difficult to detect these
peculiar heavy elements in the surface. It seems reasonable to say that the more
a TZO is evolved (e.g. WN8), the smaller is the envelope (or the distance
between the burning region and the surface), the easier it should be to observe 
signatures of a degenerate neutron-core supported burning-region. 

In fact, we can argue that WR stars have very strong winds, and all
photospheric information is obscured by the wind. This is true mainly for
hotter subtypes. Nevertheless WN8 stars are peculiar also because they have a slow wind
velocity, and hence narrow emission lines. It could be very interesting to
perform the same research of heavy elements (via echelle-spectra?) among the
WN8 population, that it is being done for RSG (see Kuchner \& Vakil, these proceedings).

The Galaxy has about 200 known WR stars (van der Hucht 2000), among which a quarter
are WNL stars (i.e. WN9 to WN6); of these a quarter are of subtype WN8.
This means that $\sim$12 TZO(WN8) could exist among the observed WR
population in the Galaxy. In the Large
Magellanic Cloud (where detailed echelle spectroscopy could also be
made) one finds 3 stars of subtype WN8 (Breysacher et al. 1999).

\section{Conclusion}

We have presented here the two known scenarii in which TZOs form following 
close binary evolution and a common envelope phase. The hydrodynamical
simulations of Terman et al. (1995) dealing with the entry of the NS inside the
envelope of a more-or-less evolved RSG allowed us to construct a paradigm
within which WN8 stars, TZOs and objects like Cyg X-3, find an
explanation. We also drew attention to the peculiarities of WN8 stars which fit
well in our "model". Finally we believe that spectroscopic observations might
be feasible to reveal the presence of a degenerate neutron core inside single WN8
stars, which would then confirm the existence of TZOs.

To the question: "Are peculiar 
Wolf-Rayet stars of type WN8 Thorne-$\dot{\mbox{Z}}$ytkow Objects?" we should now respond "Why
not?".

\acknowledgments C.F. acknowledges Kenneth Gayley for valuable
comments on the manuscript.

\end{document}